\documentclass[preprint,aps,pra,showpacs,floatfix]{revtex4}
\usepackage{graphicx}
\usepackage{times}
\usepackage{nicefrac}
\usepackage{amsmath}
\usepackage{amsfonts}
\usepackage{amssymb}
\usepackage{amsthm}
\usepackage{epsf}
\usepackage{bm}
\usepackage{bbm}

\usepackage{dcolumn}
\newcolumntype{.}{D{x}{}{-1}}

\newcommand{\be}{\begin{eqnarray}}
\newcommand{\ee}{\end{eqnarray}}
\newcommand{\la}{\langle}
\newcommand{\ra}{\rangle}

\newcommand{\eps}{\epsilon}

\newcommand{\vpar}{\partial}
\newcommand{\pr}{\prime}

\newcommand{\calE}{{\cal E}}
\newcommand{\calH}{{\cal H}}
\newcommand{\calR}{{\cal R}}
\newcommand{\bnabla}{\bm{\nabla}}

\newcommand{\beps}{\bm{\epsilon}}

\newcommand{\bmu}{\bm{\mu}}
\newcommand{\bfA}{{\bf A}}

\newcommand{\bfJ}{{\bf J}}
\newcommand{\bfL}{{\bf L}}
\newcommand{\bfP}{{\bf P}}
\newcommand{\bfR}{{\bf R}}
\newcommand{\bfS}{{\bf S}}

\newcommand{\rmd}{{\rm d}}
\newcommand{\bfk}{{\bf k}}

\newcommand{\bfp}{{\bf p}}

\newcommand{\bfs}{{\bf s}}

\newcommand{\bfr}{{\bf r}}
\newcommand{\bfx}{{\bf x}}

\newcommand{\bfxi}{{\bm{\xi}}}
\newcommand{\ddd}{\rmd^3}

\newcommand{\az}{\alpha Z}
\newcommand{\mub}{\mu_0}

\begin{document}

\title{Nuclear recoil effect on the magnetic-dipole decay rates
of atomic levels}
\author{A. V. Volotka,$^{1}$ D. A. Glazov,$^{2}$ G. Plunien,$^{1}$
V. M. Shabaev,$^{2}$ and I. I. Tupitsyn$^{2}$}

\affiliation{
$^1$ Institut f\"ur Theoretische Physik, Technische Universit\"at Dresden,
Mommsenstra{\ss}e 13, D-01062 Dresden, Germany \\
$^2$ Department of Physics, St. Petersburg State University,
Oulianovskaya 1, Petrodvorets, 198504 St. Petersburg, Russia \\
}

\begin{abstract}
The effect of finite nuclear mass on the magnetic-dipole
transition probabilities between fine-structure levels
of the same term is investigated.
Based on a rigorous QED approach a nonrelativistic formula for the recoil correction
to first order in $m_e/M$ is derived. Numerical results for
transitions of experimental interest are presented.
\end{abstract}


\maketitle

%%%%%%%%%%%%%%%%%%%%%%%%%%%%%%%%%%%%%%%%%%%%%%%%%%%%%%%%%%%%%%%%%%%%%%%%
%
%%%%%%%%%%%%%%%%%%%%%%%%%%%%%%%%%%%%%%%%%%%%%%%%%%%%%%%%%%%%%%%%%%%%%%%%
%
Investigations of transition rates and transition energies
provide direct access to most important spectroscopic properties
of atoms and ions. In addition to the detailed understanding of the
atomic structure, studies of transition rates in multicharged
ions are of great relevance in plasma diagnostics and astrophysics \cite{edlen:1984:5}.
The relative intensities of electric-dipole (E1) forbidden transitions
are frequently employed as a sensitive tool for plasma density
diagnostics and coronal lines analysis.
With the help of novel devices such as ion traps and Electron-Beam Ion-Traps
(EBITs), new experimental high-precision data for the magnetic-dipole (M1)
decays have become available \cite{traebert:2001:034501,traebert:2002:052507,
lapierre:2005:183001,lapierre:2006:052507,brenner:2007:032504}.
The M1-transition rate between the fine-structure
levels $(1s^2 2s^2 2p) \, ^2P_{3/2} \, - \, ^2P_{1/2}$
in B-like Ar$^{13+}$ ion associated with the 4412.4 \AA\, coronal line
has been measured with an accuracy better than one part per thousand
\cite{lapierre:2005:183001,lapierre:2006:052507}.
The elaborative theoretical calculation of this decay rate,
including relativistic, interelectronic-interaction, and
quantum-electrodynamic (QED) corrections, has revealed a discrepancy
between theory and experiment \cite{tupitsyn:2005:062503,volotka:2006:293}.
The total theoretical uncertainty is mainly set
by the uncalculated recoil correction together with the experimental
error of the transition energy.
In this work we present the evaluation of the recoil correction to
the M1-transition rates between fine-structure levels of the same term.

The finite nuclear mass effect on the photon emission was first taken
into consideration for E1-decays in Ref.~\cite{fried:1963:574}.
The relativistic-recoil correction of the order $(Zm_e/M)(\az)^2$
($m_e$ and $M$ are the electron and nucleus masses, respectively)
for these decays was calculated in work \cite{karshenboim:1997:4311}.
The recoil effect on the two-E1-photon transition
$2s \, - \, 1s$ and the transitions between hyperfine-structure levels
was investigated in Refs.~\cite{bacher:1984:135,shabaev:1998:907}.
Some developments for the forbidden transitions in light atoms
were presented in Refs.~\cite{pachucki:2003:012504,pachucki:2004:052502}.
However, the recoil effect on the M1-decay between fine-structure levels has
not been considered before. 
The derivation of the nonrelativistic formula for the recoil correction
to the corresponding M1-transition rates within a rigorous QED description
is the main result of the present paper.

Our consideration starts from the nonrelativistic
Hamiltonian for $N$ electrons and for the nucleus (atom) interacting
with the second-quantized radiation field.
Utilizing the transverse gauge and the Schr\"odinger representation it can be written as
(in units $\hbar=c=1$, $e=-|e|<0$)
\be
\label{H:1}
  H &=& \frac{1}{2m_e} \sum_i \left[\bfp_i^{(e)} - e \bfA (\bfr_i^{(e)})\right]^2
     + \frac{1}{2M} \left[\bfp^{(n)} + e Z \bfA (\bfr^{(n)})\right]^2
     + \sum_i V(\bfr_i^{(e)}-\bfr^{(n)})\nonumber\\
  && + \frac{1}{2} \sum_{i \ne k} \frac{\alpha}{|\bfr_i^{(e)}-\bfr_k^{(e)}|}
     + \frac{1}{2} \int \ddd x \, [\calE_t^2(\bfx) + \calH^2(\bfx)]
     + 2\mub \sum_i \bfs_i^{(e)}\cdot\calH(\bfr_i^{(e)})\,,
\ee
where the indices ``$e$'' and ``$n$'' designate the action on the electron
and nucleus variables, respectively,
$\bfA(\bfr)$ is the (transversal) vector potential of the quantized
electric ($\calE_t=-\vpar_t\bfA$) and magnetic 
($\calH=\bnabla\times\bfA$) fields,
and the operator $V(\bfr_i^{(e)}-\bfr^{(n)})$ describes
the Coulomb interaction between the $i$-th electron and the nucleus.
The last term in Eq.~(\ref{H:1}) corresponds to the interaction
of the electron spin $\bfs^{(e)}$ with the quantized radiation field,
where $\mub=|e|\hbar/2mc$ denotes the Bohr magneton.
Here we assume that the nucleus possesses zero spin.
It is most suitable to introduce relative $\bfr_i$ and center-of-mass $\bfR$ variables
for the electron-nucleus subsystem
\be
  \bfr_i &=& \bfr_i^{(e)} - \bfr^{(n)}\,,\\
  \bfR   &=& \frac{1}{M + N m_e} \left( M\bfr^{(n)} + m_e \sum_i \bfr_i^{(e)} \right)\,.
\ee
For the corresponding conjugate momenta $\bfp_i$ and $\bfP$ one derives
\be
  \bfp_i &=& \bfp_i^{(e)} - \frac{m_e}{M + N m_e} \left( \bfp^{(n)} + \sum_k \bfp_k^{(e)} \right)\,,\\
  \bfP   &=& \bfp^{(n)} + \sum_i \bfp_i^{(e)}\,.
\ee
Performing corresponding substitutions and exploiting the condition $N m_e \ll M$,
the Hamiltonian (\ref{H:1}) can be expanded up to zeroth order in $\alpha$ and to first order
in $m_e/M$. It may be written as the sum of an unperturbed ($H_0$) and
an interaction ($H_\gamma$) Hamiltonians
\be
\label{H:2}
  H &=& H_0 + H_\gamma\,,
\ee
where $H_0$ is defined as
\be
\label{H_0}
  H_0 &=& \frac{1}{2m_e} \sum_i \bfp_i^2
     + \sum_i V(\bfr_i) + \frac{1}{2} \sum_{i \ne k} \, \frac{\alpha}{|\bfr_i-\bfr_k|}
     + \frac{1}{2M}\Bigl(\sum_i \bfp_i\Bigr)^2\nonumber\\
   &&+ \frac{1}{2M} \, \bfP^2
     + \frac{1}{2} \int \ddd x \, [\calE_t^2(\bfx) + \calH^2(\bfx)]
\ee
and $H_\gamma$ describes the interaction
with the quantized radiation field $\bfA(\bfr)$,
\be
\label{H_gamma}
  H_\gamma &=& - \frac{e}{m_e} \sum_i \bfp_i\cdot\bfA(\bfR+\bfr_i)
     + 2\mub \sum_i \bfs_i^{(e)}\cdot\calH(\bfR+\bfr_i)\nonumber\\
   &&+ \frac{e}{M} \sum_{i, \, k} (\bfr_i\cdot\bnabla_\bfxi) 
       (\bfp_k\cdot\bfA(\bfxi))\Bigr|_{\bfxi=\bfR+\bfr_k}
     - 2\mub \frac{m_e}{M} \sum_{i, \, k} (\bfr_i\cdot\bnabla_\bfxi) 
       (\bfs_k^{(e)}\cdot\calH(\bfxi))\Bigr|_{\bfxi=\bfR+\bfr_k}\nonumber\\
   &&- \frac{e}{M} \sum_i \bfP\cdot\bfA(\bfR+\bfr_i)
     + \frac{eZ}{M} \, \bfP\cdot\bfA(\bfR)
     - \frac{eZ}{M} \sum_i \bfp_i\cdot\bfA(\bfR)\,.
\ee
According to the basic principles of quantum electrodynamics
\cite{berestetsky}, the $S$-matrix element corresponding
to the transition of an atom from the state $A$ to $B$ 
accompanied by the emission of the photon with wave vector $k_f$
and polarization $\eps_f$ is given by
\be
\label{S:1}
  S_{\gamma_f, B; A} &=& -i\,\int_{-\infty}^{\infty} \rmd t
  \int \rmd\bfR \int \rmd\bfr_1\dots\rmd\bfr_N\, 
  {\rm exp}\Bigl(i\frac{\bfP_B^2}{2M}t+iE_Bt+ik_f^0t-i\frac{\bfP_A^2}{2M}t-iE_At\Bigr)\nonumber\\
  &&\times\Psi^*_B(\bfR)\,\Phi^*_B(\bfr_1,\dots,\bfr_N)\,
  \la k_f,\eps_f | H_\gamma | 0 \ra\,
  \Psi_A(\bfR)\,\Phi_A(\bfr_1,\dots,\bfr_N)\,.
\ee
Here the wave function
\be
\label{Psi}  
  \Psi_A(\bfR) = e^{i\bfP_A\cdot\bfR}
\ee
describes the free translational motion
of the atomic center-of-mass
(it is supposed that the normalization volume
equals to unity).
The $N$-electron wave function
$\Phi_A(\bfr_1,\dots,\bfr_N) = \la \bfr_1,\dots,\bfr_N | A \ra$
obeys the following equation of motion
\be
\label{Phi}
  \left[ \sum_i \frac{\bfp_i^2}{2m_{\rm r}} + \sum_i V(\bfr_i)
   + \frac{1}{2} \sum_{i \ne k} \, \frac{\alpha}{|\bfr_i-\bfr_k|}
   + \sum_{i \ne k} \, \frac{\bfp_i\cdot\bfp_k}{2M}-E_A\right]
  \Phi_A(\bfr_1,\dots,\bfr_N) = 0\,.
\ee
$E_A$ and $E_B$ are the energies of initial and final
many-electron states, respectively, $m_{\rm r}=m_e M / (m_e+M)$ is the reduced mass.
To evaluate the matrix element $\la k_f,\eps_f | H_\gamma | 0 \ra$
in Eq.~(\ref{S:1})
we employ the commutation relations for the creation and
annihilation operators.
Then in the spectral expansion of the
quantized electromagnetic field only the term with wave
vector $k_f$ and polarization $\eps_f$ survives. 
In such a way, after the integration over $t$ and $\bfR$ variables,
the $S$-matrix element takes the form
\be
\label{S:2}
  S_{\gamma_f, B; A} = -i\,(2\pi)^4\,\delta^3(\bfP_B+\bfk_f-\bfP_A)\;
  \delta\Bigl(\frac{\bfP_B^2}{2M}+E_B+k_f^0-\frac{\bfP_A^2}{2M}-E_A\Bigr)
  \la B | V_\gamma | A\ra\,,
\ee
where
\be
\label{V_gamma}
  \la B | V_\gamma | A \ra &=& \int \rmd\bfr_1\dots\rmd\bfr_N\, 
  \Phi^*_B(\bfr_1,\dots,\bfr_N)\,\Biggl\{
   -\frac{e}{m_e}\sum_i\bfp_i\cdot\bfA^*_f(\bfr_i)
   + 2\mub \sum_i \bfs_i^{(e)}\cdot\calH^*_f(\bfr_i)\nonumber\\
   &&+\frac{e}{M}\sum_{i,\,k}(\bfr_i\cdot\bnabla_\bfxi) 
      (\bfp_k\cdot\bfA^*_f(\bfxi))\Bigr|_{\bfxi=\bfr_k}
   - 2\mub \frac{m_e}{M} \sum_{i, \, k} (\bfr_i\cdot\bnabla_\bfxi) 
      (\bfs_k^{(e)}\cdot\calH^*_f(\bfxi))\Bigr|_{\bfxi=\bfr_k}\nonumber\\
   &&-\frac{eZ}{M} \sum_i \bfp_i\cdot\beps^*_f\Biggr\}\,
  \Phi_A(\bfr_1,\dots,\bfr_N)\,,
\ee
$\bfA_f$ is the photon wave function,
\be
  \bfA_f(\bfr) = \frac{\beps_f\,{\rm exp}(i\bfk_f\cdot\bfr)}{\sqrt{2k^0_f}}\,,
\ee
and $\calH_f(\bfr) = \bnabla \times \bfA_f(\bfr)$.
One can notice that the last term in curly brackets of Eq.~(\ref{V_gamma}) represents
the leading recoil contribution of the order $Zm_e/M$ to the E1-decay.
In view of Eq.~(\ref{S:2}) the transition rate per unit of time
can be written as
\be
\label{W:1}
  \rmd W = \delta^3(\bfP_B+\bfk_f-\bfP_A)\;
  \delta\Bigl(\frac{\bfP_B^2}{2M}+E_B+k_f^0-\frac{\bfP_A^2}{2M}-E_A\Bigr)\,
  |\la B | V_\gamma | A\ra|^2\,\frac{\rmd\bfP_B\,\rmd\bfk_f}{(2\pi)^2}\,.
\ee
Taking into account the identity
\be
  \delta[f(x)] = \sum_n \frac{\delta(x-x_n)}{|f^\pr(x_n)|}
\ee
with $f(x_n)=0$ and assuming $\bfP_A = 0$, one can easily find 
\be
  \delta\Biggl[\frac{(k_f^0)^2}{2M}+E_B+k_f^0-E_A\Biggr] \simeq \delta(k_f^0+E_B-E_A)\,,
\ee
where we keep only physically meaningful root and neglect the contributions
of the order $(E_A-E_B)/M$ and higher.
Expanding further the photon wave function in a multipole series
(see, e.g., Refs.~\cite{tupitsyn:2005:062503,volotka:2002:1263})
we keep only the M1-component.
The first three terms in braces of Eq.~(\ref{V_gamma}) will
contribute to the M1-decay.
Finally, the M1-transition probability
between the many-electron states $| A \ra$ and $| B \ra$ accounting
for the recoil correction to first order in $m_e/M$ is given by
the following nonrelativistic formula in terms of a reduced matrix element
\be
\label{W:2}
  W = \frac{4}{3}\,\frac{\omega^3\,\mu_0^2}{2J_A+1}\,
  |\la B ||\Bigl(\bfL + 2\bfS - \frac{m_e}{M}\sum_{i,\,k}[\bfr_i\times\bfp_k]\Bigl)|| A\ra|^2\,,
\ee
where the integrations over the photon energy and angles have been carried out,
$J_A$ is the angular momentum number of the initial state,
$\bfL$ and $\bfS$ are the orbital and spin angular momentum
operators, respectively,
and $\omega \equiv k^0_f = E_A - E_B$ characterizes the transition energy.
The wave functions $|A\ra$ and $|B\ra$ are determined by Eq.~(\ref{Phi}).
However, since we restrict our consideration to the first order in $m_e/M$,
the recoil corrections in Eq.~(\ref{Phi}) can be omitted.
The M1-transition operator entering Eq.~(\ref{W:2}) coincides with
the corresponding recoil corrected magnetic moment operator
\be
  \bmu = -\mu_0\,\Bigl(\bfL + 2\bfS - \frac{m_e}{M}\sum_{i,k}[\bfr_i\times\bfp_k]\Bigr)\,,
\ee
which was derived in Ref.~\cite{phillips:1949:1803} for the evaluation of $g$ factor.
The calculation of the recoil correction to the bound-electron $g$ factor
to all orders in $\alpha Z$ can be found in
Refs.~\cite{shabaev:2001:052104,shabaev:2002:091801}.

Finally, the recoil correction to the M1-transition probability takes the form
\be
\label{W:rec}
  \Delta W^{\rm rec} = -\frac{8}{3}\,\frac{\omega^3\,\mu_0^2}{2J_A+1}\frac{m_e}{M}\;
  \la B ||\Bigl(\bfL + 2\bfS \Bigl)|| A\ra\,
  \la B ||\sum_{i,\,k}[\bfr_i\times\bfp_k]|| A\ra\,,
\ee
where $|A\ra$ and $|B\ra$ are linear combinations of the Slater
determinants constructed in terms of one-electron Schr\"odinger wave functions
and being eigenstates of the operators $\bfJ^2$, ${\rm J}_z$, $\bfL^2$, and $\bfS^2$.
This formula holds for all M1-transitions between the fine-structure
levels of the same term with arbitrary number of electrons.
An additional recoil correction to Eq.~(\ref{W:rec})
arises also from the transition energy. However, we do not
consider this contribution here, since we will further employ
the experimental value for the energy of the emitted photon.

In what follows we present some explicit results for highly charged ions
restricting to the approximation of noninteracting electrons.
Let us consider first the M1-transition between the levels
$(1s^2 2s^2 2p) \, ^2P_{3/2} \, - \, ^2P_{1/2}$
in boronlike ions. To perform the angular
integration we employ formulas presented in Refs.~\cite{artemyev:1995:1884,varshalovich}.
Accordingly, for the second reduced matrix element in Eq.~(\ref{W:rec}),
we obtain
\be
  \la (1s^2 2s^2 2p)\,^2P_{1/2} || \sum_{i,\,k}[\bfr_i\times\bfp_k] || ^2P_{3/2} \ra =
  -\frac{2}{\sqrt{3}}-\frac{4}{3\sqrt{3}}\,\sum_{c=1s,2s} \calR_{2p,c}\,
\ee
with
\be
\label{radint}
  \calR_{2p,c} = \int_0^\infty \rmd r_1\,r_1^3\,R_{2p}(r_1)R_{c}(r_1)
  \int_0^\infty \rmd r_2\,r_2^2\,R_{2p}(r_2)\frac{\rmd}{\rmd r_2} R_{c}(r_2)\,,
\ee
and $R_c$ is the nonrelativistic Schr\"odinger wave function
of an electron in state $c$.
The radial integrations in Eq.~(\ref{radint}) are performed
analytically with the following result for the recoil correction
\be
  \Delta W^{\rm rec} = \frac{8}{9} \omega^3 \mub^2\, 
    \frac{m_e}{M}\,\left(1-\frac{2^{13}}{3^{9}}\right)\,.
\ee
For the experimentally most interesting case of $^{40}$Ar$^{13+}$ ion
\cite{lapierre:2005:183001,lapierre:2006:052507}
it yields $\Delta W^{\rm rec}=0.0017$ s$^{-1}$,
utilizing the experimental value for the transition energy
$\omega = 22656.22$ cm$^{-1}$
\cite{draganic:2003:183001}.
The theoretical value for the corresponding transition energy
obtained, recently, by means of an {\it ab initio} QED approach
\cite{artemyev:2007:173004} is
in perfect agreement with the experimental one.
For this ion, the uncertainty of the recoil correction to the
transition rate
due to correlation effects is assumed to be about $20\%$
by the relative estimation as a ratio $n_c/Z$, where $n_c$
is the number of the core electrons.
Adding the recoil correction to the decay rate
calculated in Ref.~\cite{volotka:2006:293},
we obtain the total values for the transition probability
$W_{\rm total} = 104.848(1)$ s$^{-1}$ and the corresponding lifetime
$\tau_{\rm total} = 9.5376(1)$ ms. 
The comparison between our theoretical result with the most accurate
experimental value $\tau_{\rm exp} = 9.573(4)(^{+12}_{-5})$ ms (stat)(syst)
\cite{lapierre:2006:052507} reveals a remaining discrepancy of about 5.5$\sigma^-$,
where $\sigma^-$ is defined as combination of the statistical and negative
systematic uncertainties in quadrature.

Furthermore, we consider the M1-transition between the levels
$(1s^2 2s^2 2p^6 3s^2 3p) \, ^2P_{3/2} \, - \, ^2P_{1/2}$
in aluminiumlike highly charged ions.
The angular integration of the corresponding reduced matrix element yields
\be
  \la (1s^2 2s^2 2p^6 3s^2 3p)\,^2P_{1/2} || \sum_{i,\,k}[\bfr_i\times\bfp_k] || ^2P_{3/2} \ra =
  -\frac{2}{\sqrt{3}}-\frac{4}{3\sqrt{3}}\,\sum_{c=1s,2s,3s} \calR_{3p,c}\,.
\ee
After the radial integration is performed the recoil correction results in
the following expression
\be
  \Delta W^{\rm rec} = \frac{8}{9} \omega^3 \mub^2\, 
    \frac{m_e}{M}\,\left(1-\frac{3^{4}}{2^{10}}-\frac{2^{18}\, 3^{4}}{5^{11}}\right)\,.
\ee
For the case of $^{56}$Fe$^{13+}$ ion,
where the most accurate experimental result exists \cite{brenner:2007:032504},
we obtain $\Delta W^{\rm rec}=0.00057$ s$^{-1}$.
We assign a $50\%$ uncertainty for this value
due to the recoil correction beyond the one-electron approximation.
The value of the corresponding transition energy
$\omega = 18852$ cm$^{-1}$ is taken from Ref.~\cite{churilov:1993:425}.
Finally, the total theoretical values for the decay rate and the lifetime
are $W_{\rm total} = 60.44$ s$^{-1}$ and
$\tau_{\rm total} = 16.545$ ms, respectively.
These values include the interelectronic-interaction correction
with the frequency-dependent term (see, for the details
Refs.~\cite{indelicato:2004:062506,tupitsyn:2005:062503}),
the QED contribution, and the recoil correction.
The details of this calculation will be presented elsewhere.
The most accurate experimental value $\tau_{\rm exp} = 16.726^{+0.020}_{-0.010}$ ms
\cite{brenner:2007:032504} significantly deviates from our theoretical result.

Summarizing this work, the nonrelativistic
formula accounting for the recoil correction
in first order in $m_e/M$
to the M1-transition rates between the fine-structure levels
has been derived within a rigorous QED approach. Explicit results for the B-like
and Al-like highly charged ions have been obtained
within the one-electron approximation.
The values for the recoil correction together with
the total values of the transition probabilities
have been obtained for the case of B-like Ar$^{13+}$
and Al-like Fe$^{13+}$ ions.
Comparison of our total results with the most accurate experimental data
reveals remaining discrepancies, which can not be explained 
by the recoil correction.
However, we expect that further investigations of the transition rates
in ions with several electrons could elucidate the origin of this disagreement.
%
%%%%%%%%%%%%%%%%%%%%%%%%%%%%%%%%%%%%%%%%%%%%%%%%%%%%%%%%%%%%%%%%%%%%%%%%
\acknowledgments
%%%%%%%%%%%%%%%%%%%%%%%%%%%%%%%%%%%%%%%%%%%%%%%%%%%%%%%%%%%%%%%%%%%%%%%%
%
The work was supported by DFG and GSI. Further support
by RFBR (Grant No. 07-02-00126a) and INTAS-GSI (Grant No. 06-1000012-8881)
is also gratefully acknowledged.
%
%%%%%%%%%%%%%%%%%%%%%%%%%%%%%%%%%%%%%%%%%%%%%%%%%%%%%%%%%%%%%%%%%%%%%%%%
%
%%%%%%%%%%%%%%%%%%%%%%%%%%%%%%%%%%%%%%%%%%%%%%%%%%%%%%%%%%%%%%%%%%%%%%%%
%\bibliography{liter}

\begin{thebibliography}{25}
\expandafter\ifx\csname natexlab\endcsname\relax\def\natexlab#1{#1}\fi
\expandafter\ifx\csname bibnamefont\endcsname\relax
  \def\bibnamefont#1{#1}\fi
\expandafter\ifx\csname bibfnamefont\endcsname\relax
  \def\bibfnamefont#1{#1}\fi
\expandafter\ifx\csname citenamefont\endcsname\relax
  \def\citenamefont#1{#1}\fi
\expandafter\ifx\csname url\endcsname\relax
  \def\url#1{\texttt{#1}}\fi
\expandafter\ifx\csname urlprefix\endcsname\relax\def\urlprefix{URL }\fi
\providecommand{\bibinfo}[2]{#2}
\providecommand{\eprint}[2][]{\url{#2}}

\bibitem[{\citenamefont{Edl{\'e}n}(1984)}]{edlen:1984:5}
\bibinfo{author}{\bibfnamefont{B.}~\bibnamefont{Edl{\'e}n}},
  \bibinfo{journal}{Phys. Scr.} \textbf{\bibinfo{volume}{T8}},
  \bibinfo{pages}{5} (\bibinfo{year}{1984}).

\bibitem[{\citenamefont{Tr{\"a}bert et~al.}(2001)\citenamefont{Tr{\"a}bert,
  Beiersdorfer, Brown, Chen, Pinnington, and Thorn}}]{traebert:2001:034501}
\bibinfo{author}{\bibfnamefont{E.}~\bibnamefont{Tr{\"a}bert}},
  \bibinfo{author}{\bibfnamefont{P.}~\bibnamefont{Beiersdorfer}},
  \bibinfo{author}{\bibfnamefont{G.~V.} \bibnamefont{Brown}},
  \bibinfo{author}{\bibfnamefont{H.}~\bibnamefont{Chen}},
  \bibinfo{author}{\bibfnamefont{E.~H.} \bibnamefont{Pinnington}},
  \bibnamefont{and} \bibinfo{author}{\bibfnamefont{D.~B.} \bibnamefont{Thorn}},
  \bibinfo{journal}{Phys. Rev. A} \textbf{\bibinfo{volume}{64}},
  \bibinfo{pages}{034501} (\bibinfo{year}{2001}).

\bibitem[{\citenamefont{Tr{\"a}bert et~al.}(2002)\citenamefont{Tr{\"a}bert,
  Beiersdorfer, Gwinner, Pinnington, and Wolf}}]{traebert:2002:052507}
\bibinfo{author}{\bibfnamefont{E.}~\bibnamefont{Tr{\"a}bert}},
  \bibinfo{author}{\bibfnamefont{P.}~\bibnamefont{Beiersdorfer}},
  \bibinfo{author}{\bibfnamefont{G.}~\bibnamefont{Gwinner}},
  \bibinfo{author}{\bibfnamefont{E.~H.} \bibnamefont{Pinnington}},
  \bibnamefont{and} \bibinfo{author}{\bibfnamefont{A.}~\bibnamefont{Wolf}},
  \bibinfo{journal}{Phys. Rev. A} \textbf{\bibinfo{volume}{66}},
  \bibinfo{pages}{052507} (\bibinfo{year}{2002}).

\bibitem[{\citenamefont{Lapierre et~al.}(2005)\citenamefont{Lapierre,
  Jentschura, {Crespo~L\'opez-Urrutia}, Braun, Brenner, Bruhns, Fischer,
  {Gonz\'alez~Mart\'inez}, Harman, Johnson et~al.}}]{lapierre:2005:183001}
\bibinfo{author}{\bibfnamefont{A.}~\bibnamefont{Lapierre}},
  \bibinfo{author}{\bibfnamefont{U.~D.} \bibnamefont{Jentschura}},
  \bibinfo{author}{\bibfnamefont{J.~R.}
  \bibnamefont{{Crespo~L\'opez-Urrutia}}},
  \bibinfo{author}{\bibfnamefont{J.}~\bibnamefont{Braun}},
  \bibinfo{author}{\bibfnamefont{G.}~\bibnamefont{Brenner}},
  \bibinfo{author}{\bibfnamefont{H.}~\bibnamefont{Bruhns}},
  \bibinfo{author}{\bibfnamefont{D.}~\bibnamefont{Fischer}},
  \bibinfo{author}{\bibfnamefont{A.~J.} \bibnamefont{{Gonz\'alez~Mart\'inez}}},
  \bibinfo{author}{\bibfnamefont{Z.}~\bibnamefont{Harman}},
  \bibinfo{author}{\bibfnamefont{W.~R.} \bibnamefont{Johnson}},
  \bibnamefont{et~al.}, \bibinfo{journal}{Phys. Rev. Lett.}
  \textbf{\bibinfo{volume}{95}}, \bibinfo{pages}{183001}
  (\bibinfo{year}{2005}).

\bibitem[{\citenamefont{Lapierre et~al.}(2006)\citenamefont{Lapierre,
  {Crespo~L\'opez-Urrutia}, Braun, Brenner, Bruhns, Fischer,
  {Gonz\'alez~Mart\'inez}, Mironov, Osborne, Sikler
  et~al.}}]{lapierre:2006:052507}
\bibinfo{author}{\bibfnamefont{A.}~\bibnamefont{Lapierre}},
  \bibinfo{author}{\bibfnamefont{J.~R.}
  \bibnamefont{{Crespo~L\'opez-Urrutia}}},
  \bibinfo{author}{\bibfnamefont{J.}~\bibnamefont{Braun}},
  \bibinfo{author}{\bibfnamefont{G.}~\bibnamefont{Brenner}},
  \bibinfo{author}{\bibfnamefont{H.}~\bibnamefont{Bruhns}},
  \bibinfo{author}{\bibfnamefont{D.}~\bibnamefont{Fischer}},
  \bibinfo{author}{\bibfnamefont{A.~J.} \bibnamefont{{Gonz\'alez~Mart\'inez}}},
  \bibinfo{author}{\bibfnamefont{V.}~\bibnamefont{Mironov}},
  \bibinfo{author}{\bibfnamefont{C.}~\bibnamefont{Osborne}},
  \bibinfo{author}{\bibfnamefont{G.}~\bibnamefont{Sikler}},
  \bibnamefont{et~al.}, \bibinfo{journal}{Phys. Rev. A}
  \textbf{\bibinfo{volume}{73}}, \bibinfo{pages}{052507}
  (\bibinfo{year}{2006}).

\bibitem[{\citenamefont{Brenner et~al.}(2007)\citenamefont{Brenner,
  {Crespo~L\'opez-Urrutia}, Harman, Mokler, and Ullrich}}]{brenner:2007:032504}
\bibinfo{author}{\bibfnamefont{G.}~\bibnamefont{Brenner}},
  \bibinfo{author}{\bibfnamefont{J.~R.}
  \bibnamefont{{Crespo~L\'opez-Urrutia}}},
  \bibinfo{author}{\bibfnamefont{Z.}~\bibnamefont{Harman}},
  \bibinfo{author}{\bibfnamefont{P.~H.} \bibnamefont{Mokler}},
  \bibnamefont{and} \bibinfo{author}{\bibfnamefont{J.}~\bibnamefont{Ullrich}},
  \bibinfo{journal}{Phys. Rev. A} \textbf{\bibinfo{volume}{75}},
  \bibinfo{pages}{032504} (\bibinfo{year}{2007}).

\bibitem[{\citenamefont{Tupitsyn et~al.}(2005)\citenamefont{Tupitsyn, Volotka,
  Glazov, Shabaev, Plunien, {Crespo~L\'opez-Urrutia}, Lapierre, and
  Ullrich}}]{tupitsyn:2005:062503}
\bibinfo{author}{\bibfnamefont{I.~I.} \bibnamefont{Tupitsyn}},
  \bibinfo{author}{\bibfnamefont{A.~V.} \bibnamefont{Volotka}},
  \bibinfo{author}{\bibfnamefont{D.~A.} \bibnamefont{Glazov}},
  \bibinfo{author}{\bibfnamefont{V.~M.} \bibnamefont{Shabaev}},
  \bibinfo{author}{\bibfnamefont{G.}~\bibnamefont{Plunien}},
  \bibinfo{author}{\bibfnamefont{J.~R.}
  \bibnamefont{{Crespo~L\'opez-Urrutia}}},
  \bibinfo{author}{\bibfnamefont{A.}~\bibnamefont{Lapierre}}, \bibnamefont{and}
  \bibinfo{author}{\bibfnamefont{J.}~\bibnamefont{Ullrich}},
  \bibinfo{journal}{Phys. Rev. A} \textbf{\bibinfo{volume}{72}},
  \bibinfo{pages}{062503} (\bibinfo{year}{2005}).

\bibitem[{\citenamefont{Volotka et~al.}(2006)\citenamefont{Volotka, Glazov,
  Plunien, Shabaev, and Tupitsyn}}]{volotka:2006:293}
\bibinfo{author}{\bibfnamefont{A.~V.} \bibnamefont{Volotka}},
  \bibinfo{author}{\bibfnamefont{D.~A.} \bibnamefont{Glazov}},
  \bibinfo{author}{\bibfnamefont{G.}~\bibnamefont{Plunien}},
  \bibinfo{author}{\bibfnamefont{V.~M.} \bibnamefont{Shabaev}},
  \bibnamefont{and} \bibinfo{author}{\bibfnamefont{I.~I.}
  \bibnamefont{Tupitsyn}}, \bibinfo{journal}{Eur. Phys. J. D}
  \textbf{\bibinfo{volume}{38}}, \bibinfo{pages}{293} (\bibinfo{year}{2006}).

\bibitem[{\citenamefont{Fried and Martin}(1963)}]{fried:1963:574}
\bibinfo{author}{\bibfnamefont{Z.}~\bibnamefont{Fried}} \bibnamefont{and}
  \bibinfo{author}{\bibfnamefont{A.~D.} \bibnamefont{Martin}},
  \bibinfo{journal}{Nuovo Cimento} \textbf{\bibinfo{volume}{29}},
  \bibinfo{pages}{574} (\bibinfo{year}{1963}).

\bibitem[{\citenamefont{Karshenboim}(1997)}]{karshenboim:1997:4311}
\bibinfo{author}{\bibfnamefont{S.~G.} \bibnamefont{Karshenboim}},
  \bibinfo{journal}{Phys. Rev. A} \textbf{\bibinfo{volume}{56}},
  \bibinfo{pages}{4311} (\bibinfo{year}{1997}).

\bibitem[{\citenamefont{Bacher}(1984)}]{bacher:1984:135}
\bibinfo{author}{\bibfnamefont{R.}~\bibnamefont{Bacher}}, \bibinfo{journal}{Z.
  Phys. A} \textbf{\bibinfo{volume}{315}}, \bibinfo{pages}{135}
  (\bibinfo{year}{1984}).

\bibitem[{\citenamefont{Shabaev}(1998)}]{shabaev:1998:907}
\bibinfo{author}{\bibfnamefont{V.~M.} \bibnamefont{Shabaev}},
  \bibinfo{journal}{Can. J. Phys.} \textbf{\bibinfo{volume}{76}},
  \bibinfo{pages}{907} (\bibinfo{year}{1998}).

\bibitem[{\citenamefont{Pachucki}(2003)}]{pachucki:2003:012504}
\bibinfo{author}{\bibfnamefont{K.}~\bibnamefont{Pachucki}},
  \bibinfo{journal}{Phys. Rev. A} \textbf{\bibinfo{volume}{67}},
  \bibinfo{pages}{012504} (\bibinfo{year}{2003}).

\bibitem[{\citenamefont{Pachucki}(2004)}]{pachucki:2004:052502}
\bibinfo{author}{\bibfnamefont{K.}~\bibnamefont{Pachucki}},
  \bibinfo{journal}{Phys. Rev. A} \textbf{\bibinfo{volume}{69}},
  \bibinfo{pages}{052502} (\bibinfo{year}{2004}).

\bibitem[{\citenamefont{Berestetsky et~al.}(1982)\citenamefont{Berestetsky,
  Lifshitz, and Pitaevsky}}]{berestetsky}
\bibinfo{author}{\bibfnamefont{V.~B.} \bibnamefont{Berestetsky}},
  \bibinfo{author}{\bibfnamefont{E.~M.} \bibnamefont{Lifshitz}},
  \bibnamefont{and} \bibinfo{author}{\bibfnamefont{L.~P.}
  \bibnamefont{Pitaevsky}}, \emph{\bibinfo{title}{Quantum Electrodynamics}}
  (\bibinfo{publisher}{{Pergamon Press, Oxford}}, \bibinfo{year}{1982}).

\bibitem[{\citenamefont{Volotka et~al.}(2002)\citenamefont{Volotka, Shabaev,
  Plunien, Soff, and Yerokhin}}]{volotka:2002:1263}
\bibinfo{author}{\bibfnamefont{A.~V.} \bibnamefont{Volotka}},
  \bibinfo{author}{\bibfnamefont{V.~M.} \bibnamefont{Shabaev}},
  \bibinfo{author}{\bibfnamefont{G.}~\bibnamefont{Plunien}},
  \bibinfo{author}{\bibfnamefont{G.}~\bibnamefont{Soff}}, \bibnamefont{and}
  \bibinfo{author}{\bibfnamefont{V.~A.} \bibnamefont{Yerokhin}},
  \bibinfo{journal}{Can. J. Phys.} \textbf{\bibinfo{volume}{80}},
  \bibinfo{pages}{1263} (\bibinfo{year}{2002}).

\bibitem[{\citenamefont{Phillips}(1949)}]{phillips:1949:1803}
\bibinfo{author}{\bibfnamefont{M.}~\bibnamefont{Phillips}},
  \bibinfo{journal}{Phys. Rev.} \textbf{\bibinfo{volume}{76}},
  \bibinfo{pages}{1803} (\bibinfo{year}{1949}).

\bibitem[{\citenamefont{Shabaev}(2001)}]{shabaev:2001:052104}
\bibinfo{author}{\bibfnamefont{V.~M.} \bibnamefont{Shabaev}},
  \bibinfo{journal}{Phys. Rev. A} \textbf{\bibinfo{volume}{64}},
  \bibinfo{pages}{052104} (\bibinfo{year}{2001}).

\bibitem[{\citenamefont{Shabaev and Yerokhin}(2002)}]{shabaev:2002:091801}
\bibinfo{author}{\bibfnamefont{V.~M.} \bibnamefont{Shabaev}} \bibnamefont{and}
  \bibinfo{author}{\bibfnamefont{V.~A.} \bibnamefont{Yerokhin}},
  \bibinfo{journal}{Phys. Rev. Lett.} \textbf{\bibinfo{volume}{88}},
  \bibinfo{pages}{091801} (\bibinfo{year}{2002}).

\bibitem[{\citenamefont{Artemyev et~al.}(1995)\citenamefont{Artemyev, Shabaev,
  and Yerokhin}}]{artemyev:1995:1884}
\bibinfo{author}{\bibfnamefont{A.~N.} \bibnamefont{Artemyev}},
  \bibinfo{author}{\bibfnamefont{V.~M.} \bibnamefont{Shabaev}},
  \bibnamefont{and} \bibinfo{author}{\bibfnamefont{V.~A.}
  \bibnamefont{Yerokhin}}, \bibinfo{journal}{Phys. Rev. A}
  \textbf{\bibinfo{volume}{52}}, \bibinfo{pages}{1884} (\bibinfo{year}{1995}).

\bibitem[{\citenamefont{Varshalovich et~al.}(1988)\citenamefont{Varshalovich,
  Moskalev, and Khersonskii}}]{varshalovich}
\bibinfo{author}{\bibfnamefont{D.~A.} \bibnamefont{Varshalovich}},
  \bibinfo{author}{\bibfnamefont{A.~N.} \bibnamefont{Moskalev}},
  \bibnamefont{and} \bibinfo{author}{\bibfnamefont{V.~K.}
  \bibnamefont{Khersonskii}}, \emph{\bibinfo{title}{Quantum Theory of Angular
  Momentum}} (\bibinfo{publisher}{{World Scientific, Singapore}},
  \bibinfo{year}{1988}).

\bibitem[{\citenamefont{Dragani{\'c} et~al.}(2003)\citenamefont{Dragani{\'c},
  {Crespo~L\'opez-Urrutia}, DuBois, Fritzsche, Shabaev, {Soria~Orts}, Tupitsyn,
  Zou, and Ullrich}}]{draganic:2003:183001}
\bibinfo{author}{\bibfnamefont{I.}~\bibnamefont{Dragani{\'c}}},
  \bibinfo{author}{\bibfnamefont{J.~R.}
  \bibnamefont{{Crespo~L\'opez-Urrutia}}},
  \bibinfo{author}{\bibfnamefont{R.}~\bibnamefont{DuBois}},
  \bibinfo{author}{\bibfnamefont{S.}~\bibnamefont{Fritzsche}},
  \bibinfo{author}{\bibfnamefont{V.~M.} \bibnamefont{Shabaev}},
  \bibinfo{author}{\bibfnamefont{R.}~\bibnamefont{{Soria~Orts}}},
  \bibinfo{author}{\bibfnamefont{I.~I.} \bibnamefont{Tupitsyn}},
  \bibinfo{author}{\bibfnamefont{Y.}~\bibnamefont{Zou}}, \bibnamefont{and}
  \bibinfo{author}{\bibfnamefont{J.}~\bibnamefont{Ullrich}},
  \bibinfo{journal}{Phys. Rev. Lett.} \textbf{\bibinfo{volume}{91}},
  \bibinfo{pages}{183001} (\bibinfo{year}{2003}).

\bibitem[{\citenamefont{Artemyev et~al.}(2007)\citenamefont{Artemyev, Shabaev,
  Tupitsyn, Plunien, and Yerokhin}}]{artemyev:2007:173004}
\bibinfo{author}{\bibfnamefont{A.~N.} \bibnamefont{Artemyev}},
  \bibinfo{author}{\bibfnamefont{V.~M.} \bibnamefont{Shabaev}},
  \bibinfo{author}{\bibfnamefont{I.~I.} \bibnamefont{Tupitsyn}},
  \bibinfo{author}{\bibfnamefont{G.}~\bibnamefont{Plunien}}, \bibnamefont{and}
  \bibinfo{author}{\bibfnamefont{V.~A.} \bibnamefont{Yerokhin}},
  \bibinfo{journal}{Phys. Rev. Lett.} \textbf{\bibinfo{volume}{98}},
  \bibinfo{pages}{173004} (\bibinfo{year}{2007}).

\bibitem[{\citenamefont{Churilov and Levashov}(1993)}]{churilov:1993:425}
\bibinfo{author}{\bibfnamefont{S.~S.} \bibnamefont{Churilov}} \bibnamefont{and}
  \bibinfo{author}{\bibfnamefont{V.~E.} \bibnamefont{Levashov}},
  \bibinfo{journal}{Phys. Scr.} \textbf{\bibinfo{volume}{48}},
  \bibinfo{pages}{425} (\bibinfo{year}{1993}).

\bibitem[{\citenamefont{Indelicato et~al.}(2004)\citenamefont{Indelicato,
  Shabaev, and Volotka}}]{indelicato:2004:062506}
\bibinfo{author}{\bibfnamefont{P.}~\bibnamefont{Indelicato}},
  \bibinfo{author}{\bibfnamefont{V.~M.} \bibnamefont{Shabaev}},
  \bibnamefont{and} \bibinfo{author}{\bibfnamefont{A.~V.}
  \bibnamefont{Volotka}}, \bibinfo{journal}{Phys. Rev. A}
  \textbf{\bibinfo{volume}{69}}, \bibinfo{pages}{062506}
  (\bibinfo{year}{2004}).

\end{thebibliography}

%%%%%%%%%%%%%%%%%%%%%%%%%%%%%%%%%%%%%%%%%%%%%%%%%%%%%%%%%%%%%%%%%%%%%%%%
%
%%%%%%%%%%%%%%%%%%%%%%%%%%%%%%%%%%%%%%%%%%%%%%%%%%%%%%%%%%%%%%%%%%%%%%%%
\end{document}